%% file: WCNC_2017.tex
\documentclass[letterpaper, onecolumn,margin=0.5in]{IEEEtran}
\normalsize
\usepackage{amsmath,amsfonts,amssymb,amsbsy,url,verbatim}
\usepackage{comment}
\usepackage{graphicx}
\usepackage{url}
\usepackage{balance}
\usepackage{times}
\usepackage{dsfont}
\usepackage{subfigure}
\usepackage{cite}
\usepackage{bm}
\usepackage{epstopdf}
\usepackage{algorithm,algorithmic, color}
\usepackage{amsthm}
\usepackage{bm}
\usepackage{tikz}
\usetikzlibrary{shapes,arrows,intersections}
\usetikzlibrary{patterns,decorations}
\usetikzlibrary{mindmap}
\usetikzlibrary{calc}
\usepackage{ellipsis}
\usetikzlibrary{decorations.pathreplacing,decorations.markings,shapes.geometric,arrows,decorations.pathmorphing,decorations.footprints,fadings,calc,trees,mindmap,shadows,decorations.text,patterns,positioning,shapes,matrix,fit,chains}

\newcommand{\mat}[1]{\mathbf{#1}}

\newcommand{\bi}{\begin{itemize}}
\newcommand{\ei}{\end{itemize}}
\newcommand{\ben}{\begin{enumerate}}
\newcommand{\een}{\end{enumerate}}
\newcommand{\bc}{\begin{cases}}
\newcommand{\ec}{\end{cases}}
\newcommand{\bd}{\begin{description}}
\newcommand{\ed}{\end{description}}

\newcommand{\be}{\begin{equation}}
\newcommand{\ee}{\end{equation}}
\newcommand{\bea}{\begin{eqnarray}}
\newcommand{\eea}{\end{eqnarray}}

\IEEEoverridecommandlockouts
\input{BlocksStyle}
\input{SumNodesStyle}

\input{HighlightingStyle}
\input{LinesStyle}

\begin{document}

\title{Low-complexity Receiver for Multi-Level \\Polar Coded Modulation in \\Non-Orthogonal Multiple Access}

\author{Beatrice~Tomasi, Fr\'ed\'eric~Gabry, Valerio~Bioglio, Ingmar~Land, Jean-Claude~Belfiore\\
Mathematical and Algorithmic Sciences Laboratory, Huawei Technologies Co. Ltd.\\
France Research Center, 20 quai du Point du Jour, 92100 Boulogne Billancourt, France\\
 \small{\texttt{\{beatrice.tomasi, frederic.gabry, valerio.bioglio, ingmar.land, jean.claude.belfiore\}@huawei.com}}}

\maketitle

\begin{abstract}
Non-orthogonal multiple access (NOMA) schemes have been proved to increase the multiple-access achievable rate with respect to orthogonal multiple access (OMA). In this paper we propose a novel communication system that combines multi-level coded modulation and polar codes in a NOMA scenario. Computational complexity decreases with the proposed scheme with respect to state-of-the-art solutions. We also highlight the trade-off between error rate performance and computational complexity. 
\end{abstract}


\section{Introduction}

\input{intro.tex}

\section{Communication system}
\label{sec:II}

In this section we present the communication system considered in this paper, namely: $i$) the transmitter side consisting of a MLCM scheme with code rate design, $ii$) a non-orthogonal multiple access scheme, where multiple users are allowed to transmit simultaneously information over the same resources, and $iii$) the receiver structure, consisting in multi-user detection and multi-stage decoding.

\subsection{Transmitter Side: Multi-level Coded Modulation}
\input{fig_trans.tex}
\input{multi_level.tex}

\subsection{Code Rate Design for MLCM}\label{sec:code_rate_design}

\input{code_rate_design.tex}

\subsection{Non-Orthogonal Multiple Access}
\begin{figure}[!ht]
  \centering
\resizebox{0.45\textwidth}{!}{ \begin{tikzpicture}

\node[scale=1.5,circle,draw]  at (-3,-3) (VN1) {};
\node[left=0.2cm of VN1] {VN, $d_{\text{VN}} = 2$};
\node[scale=1.5,circle,draw]  at (-1,-3) (VN2) {};
\node[scale=1.5,circle,draw]  at (1,-3) (VN3) {};
\node[scale=1.5,circle,draw]  at (3,-3) (VN4) {};
\node[scale=1.5,circle,draw]  at (5,-3) (VN5) {};
\node[scale=1.5,circle,draw]  at (7,-3) (VN6) {};
\node[BlocksStyle]  at (-1,-6) (FN1) {};
\node[left=0.2cm of FN1,text=black] {FN, $M=4$, $d_{\text{FN}} = U = 3$};
\node[BlocksStyle]  at (1,-6) (FN2) {};
\node[BlocksStyle]  at (3,-6) (FN3) {};
\node[BlocksStyle]  at (5,-6) (FN4) {};
\path (VN1) edge[-,black]  (FN2);
\path (VN1) edge[-,black]  (FN4);
\path (VN2) edge[-,black]  (FN1);
\path (VN2) edge[-,black]  (FN3);
\path (VN3) edge[-,black]  (FN1);
\path (VN3) edge[-,black]  (FN2);
\path (VN4) edge[-,black]  (FN3);
\path (VN4) edge[-,black]  (FN4);
\path (VN5) edge[-,black]  (FN1);
\path (VN5) edge[-,black]  (FN4);
\path (VN6) edge[-,black]  (FN2);
\path (VN6) edge[-,black]  (FN3);
\end{tikzpicture}}
 \caption{Graph between users and sub-carriers.}
  \label{fig:graph1}
\end{figure}

In this paper, we use as non-orthogonal multiple access scheme the so-called SCMA scheme, first proposed in~\cite{SCMA}, where a sparse rank deficient user-resources allocation matrix is used. The main idea of SCMA consists in overloading the sub-carriers in a controlled way. This can be represented as in the graph of Fig.~\ref{fig:graph1} where users are variable nodes (VNs), and sub-carriers are function nodes (FNs). Thus the graph indicates the structure of the interference. 

The proposed communication system does not exploit a specific design of the codebook between constellation symbols and sub-carriers, and we assume that the symbols are repeated across different sub-carriers. For these reasons, the proposed system can be used in any NOMA scheme. The received signal, $\mat{Y} \in {\mathbb{C}^{M\times N}}$ can be expressed as:
\begin{equation}
\mat{Y} = \mat{H} \mat{X}+\mat{W},
\end{equation}
where $\mat{H} \in \mathbb{C}^{M\times N_u}$, is the matrix of channel coefficients over $M$ sub-carriers between the receiver and $N_u$ users. The channel coefficients are assumed to be independent and identically distributed (i.i.d.) drawn from a complex Gaussian distribution, $h_{m,k} \sim \mathcal{CN}(0,1)$, $\forall k$, $\forall m$ and the noise coefficients are drawn as $w_{m,k} \sim \mathcal{CN}(0,\sigma^{2})$. The transmitted symbols from all users over all sub-carriers are contained in $\mat{X} \in \mathbb{C}^{N_u \times N}$.

\subsection{Receiver side: Message Passing Algorithm and Multi-Stage Decoding}

\input{fig_rec.tex} 

Fig.~\ref{fig:rx_side} shows the structure of the receiver for multi-user detection and multi-stage decoding. 
We first detail the important aspects related to MU detection. At stage $l$, the codewords decoded in previous stages, $\mat{c}_i$, for $i = 0,\cdots, l-1$, for all users are assumed to be correctly decoded and canceled from the received vector $\mat{y}$. Then, the Log Likelihood Ratios (LLRs) of the codebits of the codewords at stage $l$ are computed by the Message Passing Algorithm (MPA) \cite{MPA}. The LLR at stage $l$, symbol time $t$, and user $u$, is expressed as $\text{LLR}_l(u,t) = \log{\Big(\frac{P[y_f(t)|c_l(u,t) = 0]}{P[y_f(t)|c_l(u,t) = 1]}\Big)}$, where $y_f(t)$ is the received signal at symbol time $t$ over the $f$-th sub-carrier. The MPA takes into account the users and sub-carrier allocation graph. In the following we express the function node update since it represents the most-contributing step to the overall MPA complexity. For the proposed MLCM scheme, the function node update for user $\tilde{u}$, at decoding stage $l$, given the decoded bits at previous stages $\mat{c}_i$, for $i = 0, \cdots, l-1$, can be written as:

\begin{eqnarray}
\tiny
{\mat{q}}_f(c_l(\tilde{u}))&=&\!\!\!\!\!\!\!\!\!\!\!\!\!\!\!\! \underset{\mat{c}_l(\mat{u}) \in \{0,1\}^{U-1}}{ \sum}\!\!\!\!\!\!\!\!\!\!\!\mat{p}(\mat{c}_l(\mat{u}))\mat{p}(y_f|\mat{c}_l(\mat{u}), c_l(\tilde{u})) \nonumber \\
&=& \alpha \underset{\mat{x}(\mat{c}_l)\in \mathcal{S}_l(\mat{c}_l, \mat{c}_i)}{\sum}\mat{p}(\mat{c}_l(\mat{u})) \nonumber \\
&&\!\!\!\!\!\!\!\!\!\!\!\!\!\!\!\!\!\!\!\!\!\!\!\!\!\!\!\!\!\!\!\!\!\!\!\!\!\!\!\!\!\!\!\!\!\!\!\!\!\!\underset{x(c_l)\in \mathcal{S}_l(c_l, c_i)}{\sum}\!\!\!\!\!\!\!\!\!\!\!\exp{\Big(-\frac{|y_f-\mat{H}(f,[\tilde{u},\mat{u}])[x(c_l), \mat{x}(\mat{c}_l)]^T|^2}{2\sigma^2}\Big)}\label{eq:function_node_update_MLCM},
\end{eqnarray}
where $\alpha = 1/(\sqrt{2\pi}\sigma)$, $\mathcal{S}_l$ represents the resulting symbol constellation at stage $l$, and $\mat{c}_l(\mat{u})$ corresponds to the vector of possible realizations for the $l$-th bit sent by all users other than $\tilde{u}$. The row vector $\mat{H}(f,[\tilde{u},\mat{u}])$ contains the channel coefficients corresponding to sub-carrier $f$ and all users transmitting simultaneously with user $\tilde{u}$.  The sequence of LLRs is then decoded by the polar decoder at level $l$ through CRC-aided list decoding. For all users, the information bits are then re-encoded and fed in the MPA at the next level, $l+1$, in order to select the corresponding constellation.

In the system with BICM combined with turbo codes, the MPA has function node update computed as:
\begin{eqnarray}
\tiny
	{\mat{q}}_f(x(\tilde{u})) \!&=& \!\!\!\!\underset{\mat{x}(\mat{u})}{ \sum}\mat{p}(\mat{x}(\mat{u}))\mat{p}(y_f|\mat{x}(\mat{u}), x(\tilde{u})) \nonumber \\
	&=&\alpha \underset{\mat{x}(\mat{u})}{ \sum}\mat{p}(\mat{x}(\mat{u})) \nonumber \\
	&&\!\!\!\!\!\!\!\!\!\!\!\!\!\!\!\!\!\!\!\!\!\!\!\!\!\!\!\!\exp{\Big(-\frac{|y_f-\mat{H}(f,[\tilde{u},\mat{u}])[x(\tilde{u}), \mat{x}(\mat{u})]^T|^2}{2\sigma^2}\Big)}.\label{eq:function_node_update_BICM}
\end{eqnarray}
where $\mat{x}(\mat{u})$ is the vector of possible constellation symbols sent by the other users transmitting simultaneously with $\tilde{u}$. These length-$N$ streams of LLRs are then fed to the turbo decoder.

\section{Numerical results}\label{section_simulations}

This section contains a numerical evaluation and comparison of the proposed communication scheme with that of the state of the art consisting of BICM with turbo codes. In this section, we particularly focus on the computational complexity and the error rate performance, and we discuss the trade-off between the two performance measures.  

\subsection{Scenario}

We consider the scenario depicted in Fig.~\ref{fig:graph1} where $N_u= 6$ users transmit over $M = 4$ sub-carriers to a single-antenna device. Only a subset of these users, of cardinality $U = 3$, transmit simultaneously data over each of the sub-carriers. Every user transmits $K=512$ information bits. We consider a MLCM system with $L = 4$ levels and codeword length over every level of $N_{\text{polar}} = 256$ bits. For the case of BICM with turbo codes, the codeword length is $N_{\text{turbo}}= 1024$ bits. 

\subsection{Computational Complexity and Discussion}

\begin{figure}[thb]
\centering
		\includegraphics[scale =0.6]{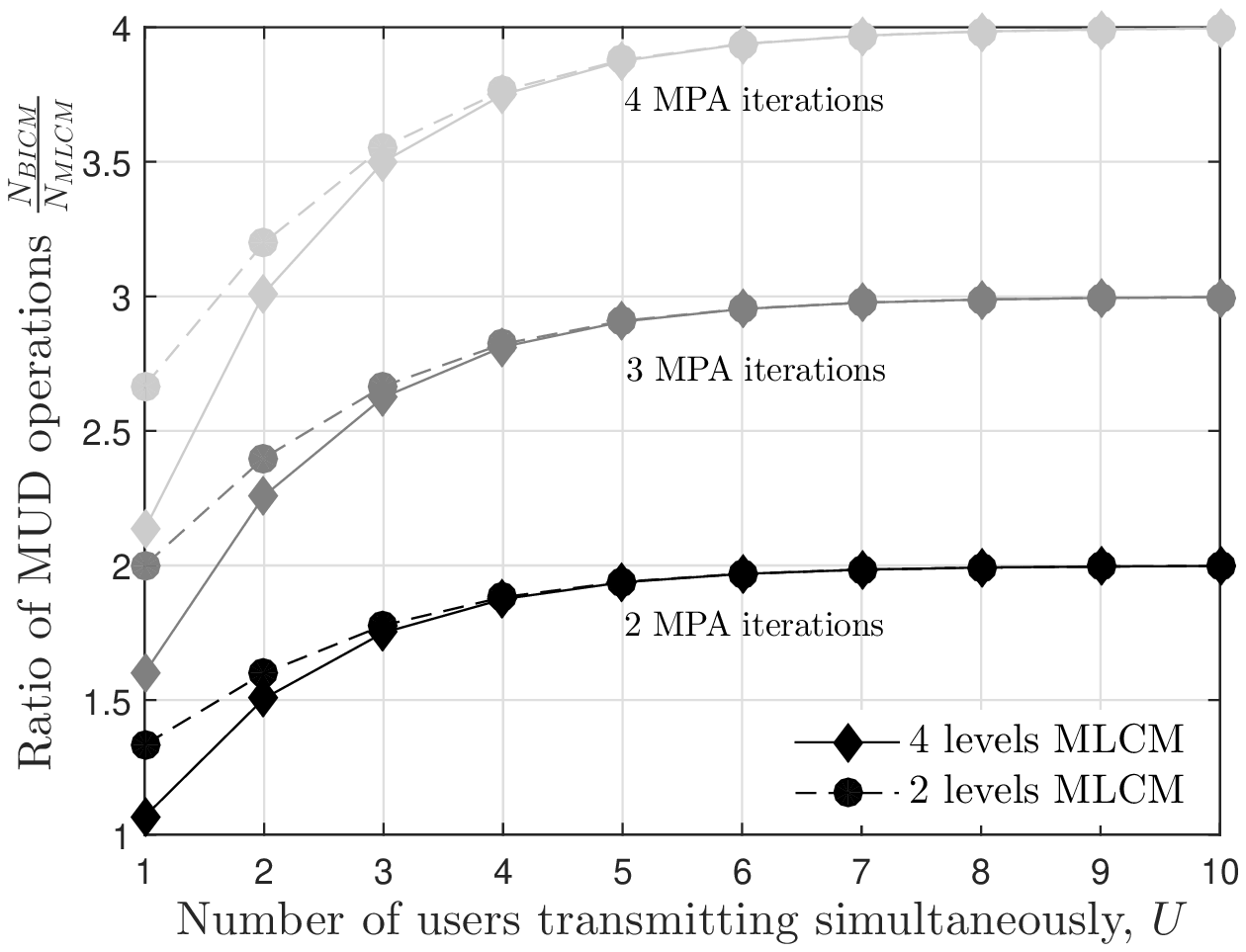}\caption{Ratio between number of FLOPs in BICM-turbo and MLCM-polar per FN during MU detection as a function of number of users transmitting over the same sub-carrier.}\label{fig:complexity}
\end{figure}

From equations~\eqref{eq:function_node_update_MLCM} and~\eqref{eq:function_node_update_BICM}, we can derive the number of FLoating Point Operations (FLOPs) required for MU-detection per sub-carrier in the proposed scheme and in the state-of-the-art communication system. In the case of BICM with turbo codes, the number of FLOPs per sub-carrier is:
\begin{equation*}
N_{\text{BICM}} = I_{\text{MPA}} 2^{LU},
\end{equation*}
where $I_{\text{MPA}}$ is the number of iterations of MPA.

In the proposed scheme, the FN update is performed at every decoding stage. The number of FLOPs per sub-carrier and per stage is equal to $2^{(L-l)U}$, where $2^{L-l}$ is the size of the remaining constellation at stage $l$. Therefore, the total number of FLOPs for MLCM is:
\begin{equation*}
N_{\text{MLCM}} = \sum_{l = 0}^{L-1} 2^{(L-l)U} = 2^U \frac{2^{UL}-1}{2^U-1}.
\end{equation*} 
Note that the MPA does not iterate for MLCM since the bit-symbol mapping used for MLCM makes users indistinguishable at the bit level, thus making MPA iterations not as effective as for the BICM case. 
Fig.~\ref{fig:complexity} shows the ratio $\frac{N_{\text{BICM}}}{N_{\text{MLCM}}}$ as a function of $U$. Dashed and solid lines correspond to a $4$-QAM constellation, i.e., $L= 2$, and $16$-QAM constellation, $L = 4$, respectively. 

\subsection{Performance Comparison between MLCM with Polar Codes and BICM with Turbo Codes}

The rate profile that we used for the analyzed scheme has the following number of information bits $K_0 = 9$, $K_1 = 70$, $K_2 = 185$, and $K_3 = 248$. 
We also apply a length-8 CRC at every level, as it improves the decoding performance of polar codes, as shown in~\cite{CA_polar}. 
This means that for example at level $0$ only $K_0-8=1$ information bit is transmitted, and hence we choose to suppress the first level and transmit data across the last three levels. This is possible by fixing $b_0 = 0$ for all users, and using such information at the decoder. 
In this way, the amount of operations performed for MU detection at the function node is  $N_{\text{MLCM}} = \sum_{l = 1}^{L-1} 2^{3(L-l)} = 584$ against the $2^{UL} = 4096$ performed by the MU detection for the state of the art. 

\begin{figure}[thb]
\centering
		\includegraphics[width=0.5\textwidth]{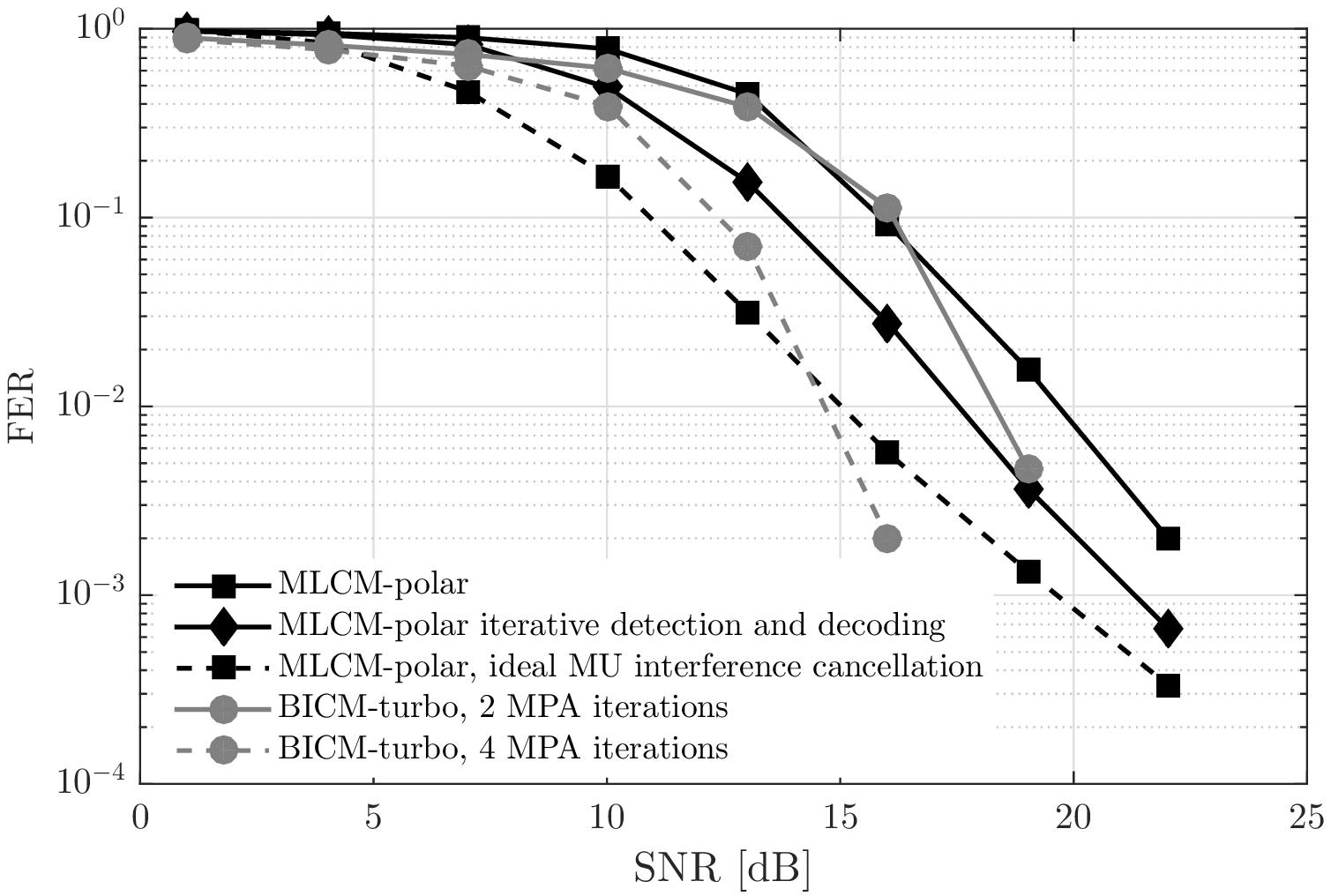}\caption{FER as a function of SNR for MLCM with polar coding and the state-of-the-art system with BICM and turbo coding.}\label{fig:performance}
\end{figure}

Fig.~\ref{fig:performance} shows the average user frame error rate as a function of the SNR. Black curves indicate the performance of the proposed scheme, while gray curves are obtained with BICM and turbo coding. These latter are obtained with $10$ turbo decoder iterations and they differ from each other in the number of MPA iterations. The proposed scheme is also affected by multi-user interference. Increasing the number of iterations in the MPA scheme does not bring any beneficial effect in performance.

The error rate curve indicated as MLCM-polar approaches the curve representing the error rate performance for BICM-turbo with $2$ MPA iterations. However, we highlight that for the MLCM-polar scheme MU detection requires $584$ FLOPs against the $8192$ FLOPs performed at each FN in BICM-turbo with $I_{\text{MPA}} = 2$. 
The dashed gray curve represents the error rate achieved when MPA converges, i.e., after $4$ iterations. We compare this latter with the dashed black curve with square markers that shows the error rate performance for MLCM-polar when at every stage the MU detector knows perfectly the bits transmitted by the interfering users. In this case the proposed scheme outperforms the state-of-the-art up to SNR equal to $14$~dB. 
For larger SNR, the BICM-turbo curve with $4$ MPA iterations outperforms MLCM-polar since the codewords of the two schemes have very different lengths, namely $N_{\text{turbo}}=1024$ and $N_{\text{polar}}=256$. However, we recall that the computational complexities per FN for MLCM-polar and BICM-turbo are $584$ and $16385$, respectively.

The dashed curve indicated with diamond markers shows the error rate performance  obtained with MLCM-polar when, at each stage, MU detection and decoding are iterated until either the CRC is successful or a fixed maximum number of iterations is achieved. 
Even though FER decreases, it does not reach the lower bound due to slow convergence and the fixed maximum number of iterations\footnote{In this case it was set to $4$.}. The computational complexity in this case increases with the average number of iterations per stage, however since data are transmitted over the last $3$ layers, it is still smaller than that for BICM-turbo with $4$ MPA iterations.

\section{Conclusions}
\label{sec:IV}
In this work, we proposed a novel scheme that is flexible in its implementation where multi-level polar coded modulation is used to support a NOMA scheme with smaller computational complexity than current state-of-the-art solutions. 
Numerical results suggest the need for a better MU detection algorithm that exploits the structure of the transmitted signal. We observed that our solution brings significant computational complexity gains compared to state-of-the-art solutions and we highlighted the trade-offs between complexity and performance in our NOMA scenario.

\bibliographystyle{IEEEtran}
\bibliography{IEEEabrv,refen2}
\end{document}

%% file: BlocksStyle.tex
\tikzset
{
	BlocksStyle/.style =
	{
		shape			= rectangle,			
		rounded corners	= 0.0cm,				
		minimum height	= 0.7cm,				
		minimum width	= 0.9cm,				
		rotate			= 0,					
		scale			= 1.0,					
		draw			= black,				
		line width		= 0.00cm,				
		align			= center,				
		text			= black,				
		font			= \normalsize\normalfont,	
		inner xsep		= 0.2cm,				
		inner ysep		= 0.2cm,				
	}
}

\tikzset
{
	BlocksStyleb/.style =
	{
		shape			= rectangle,			
		rounded corners	= 0.0cm,				
		minimum height	= 0.7cm,				
		minimum width	= 0.9cm,				
		rotate			= 0,					
		scale			= 1.0,					
		draw			= black,				
		line width		= 0.02cm,				
		fill			= white,				%
		align			= center,				
		text			= black,				
		font			= \Largesize\normalfont,	
		inner xsep		= 0.2cm,				
		inner ysep		= 0.2cm,				
	}
}

\tikzset
{
	WideBlocksStyle/.style =
	{
		BlocksStyle,
		text width		= 2.0cm,				
	}
}

%% file: SumNodesStyle.tex
\tikzset
{
	SumNodesStyle/.style =
	{
		shape			= circle,				
		minimum size	= 0.5cm,				
		rotate			= 0,					
		scale			= 0.6,					
		draw			= black,				
		line width		= 0.02cm,				
		align			= center,				
		text			= black,				
		inner sep=0pt   ,
 		text width		= 1.2cm,				
	}
}

%% file: HighlightingStyle.tex
\tikzset
{
	HighlightingStyle/.style =
	{
		color				= blue,	
		line width			= 0.04cm,			
		arrows				= -,				
		dotted,
	}
}	

\tikzset
{
	HighlightingStyleD/.style =
	{
		color				= blue,	
		line width			= 0.04cm,			
		arrows				= -,				
		dashed,
	}
}	

\tikzset
{
	HighlightingStyleB/.style =
	{
		color				= black,	
		line width			= 0.04cm,			
		arrows				= -,				
		solid,
	}
}

\tikzset
{
	HighlightingStyleE/.style =
	{
		color				= red,	
		line width			= 0.05cm,			
		arrows				= -,				
		dashed,
	}
}	
\tikzset
{
	HighlightingStyleC/.style =
	{
		color				= black,
		fill=white,
		text=black,	
		line width			= 0.05cm,			
		arrows				= -,				
 		rounded corners		= 0.0cm,			%
		solid,
	}
}

%% file: LinesStyle.tex
\tikzset
{
	LinesStyle/.style =
	{
		color				= black,	
		line width			= 0.02cm,			
		solid,
	}
}	

\tikzset
{
	LinesStyleC/.style =
	{
		color				= black,	
		line width			= 0.08cm,			
		solid,
	}
}

\tikzset
{
	LinesStyleR/.style =
	{
		color				= black,	
		line width			= 0.08cm,			
		densely dotted,
	}
}

\tikzset
{
	LinesStyleE/.style =
	{
		color				= red,	
		line width			= 0.02cm,			
		arrows				= -latex',			
		solid,
	}
}

\tikzset
{
	LinesStyleb/.style =
	{
		color				= black,	
		line width			= 0.02cm,			
		arrows				= -latex',			
 		shorten <			= 0.0cm,			
		solid,
	}
}

%% file: intro.tex
Fifth Generation (5G) cellular communication networks will have to support larger numbers of connected devices~\cite{cisco}. Multiple access techniques, currently in use and standardized in the current 3GPP-LTE, are based on orthogonal division of resources, e.g., Orthogonal Frequency-Division Multiple Access (OFDMA). As opposed to OMA, the concept of NOMA has arisen as a potential enabler for massive connectivity, improved spectral efficiency, and low transmission latency and signaling cost, which are critical aspects in the envisioned scenarios relevant to future 5G networks~\cite{NOMA,SCMA,andrews5G,5GTutorial,NOMA_Magazine}.

Although NOMA techniques are very promising due to their potential considerable performance gains, there exist several challenges towards their implementation in 5G systems as pointed out in, e.g., \cite{NOMA_Magazine}. Complexity among system scalability, signaling design that supports (possibly massive) MIMO, and CSI acquisition are some of the challenges that are critical to the implementation of NOMA systems. In this paper we focus on the aspect that might potentially be the main shortcoming of NOMA techniques, namely its complexity. In particular, receivers for NOMA \cite{NOMA_Magazine} schemes are characterized by an increased computational complexity compared to OMA due to the higher multi-user interference. The NOMA schemes proposed in the literature, e.g., in~\cite{NOMA,SCMA} provide a structure to the interference, which can then be exploited at the receiver side in order to contain the computational complexity. An example is Sparse Coded Multiple Access (SCMA)~\cite{SCMA}, which was recently proposed in the literature as a low-complexity implementation of NOMA. The main idea of SCMA is to exploit the sparsity of the allocation matrix  between the sub-carriers and users in order to reduce the computational complexity in multi-user detection; however as shown in this work, keeping the matrix sparse may not be sufficient to contain the receiver complexity.

In this paper, we propose and evaluate a novel communication architecture that has lower computational complexity at the receiver than state-of-the-art solutions. 
The proposed scheme consists in the novel combination of several techniques at different levels/stages of the communication scheme: polar codes~\cite{polar}, multi-level coded modulation (MLCM) with multi-stage decoding (MSD)~\cite{mlc99}, and NOMA. 
Initially constructed to achieve capacity over binary-input discrete memoryless channels, polar codes were shown to achieve remarkable error-rate performance at finite lengths, \cite{CA_polar,list_decoding}, outperforming state-of-the-art channel codes such as turbo codes and Low Density Parity Check (LDPC) codes. As highlighted in \cite{pcm13}, polar codes are furthermore particularly adapted to MLCM since multi-level polar codes together with MSD and successive cancellation list (SCL) decoding achieve the coded modulation capacity for arbitrary $M$-ary signal constellations in case of a memoryless transmission channel.  
In this work we propose a novel use of these techniques by constructing a polar-coded multi-level modulation scheme specifically adapted to a non-orthogonal multiple-access scheme, which is of considerable importance for practical future networks scenarios. 
We compare our scheme to the currently used 3GPP-LTE solution, i.e., Bit-Interleaved Coded Modulation (BICM) with turbo codes in terms of error rate performance and computational complexity.
The considered scenario of multiple devices having to simultaneously communicate to a single antenna device represents, the case where, for example, machines communicate with each other directly~\cite{M2M}, or users are connected to a small cell base station, where antenna elements may be few. The proposed communication architecture is novel for the NOMA scenario and its implementation is flexible in terms of rate design and multi-user (MU) detection algorithm that can be used in the system. Moreover, we show that we significantly outperform the state of the art solution in terms of the main shortcoming of current MU detection schemes, i.e., computational complexity.

This paper is organized as follows. In Section \ref{sec:II} we describe thoroughly the proposed communication scheme, i.e., our transmitter, the multiple-access scheme, and our receiver structure. In Section \ref{section_simulations} we investigate the performance of our solution from both performance and complexity perspectives. Finally Section \ref{sec:IV} concludes this paper.

%% file: fig_trans.tex
\begin{figure}[!htb]

\begin{center}
\resizebox{0.5\textwidth}{!}{
\begin{tikzpicture}
[
	xscale	= 1,	
	yscale	= 1,	
]

\matrix
[ampersand replacement=\&,
	row sep		= 0.7cm,
	column sep	= 1.09cm,
]
{
   \node (bits) {}; \&[-3ex]
	\node (nInput)  [draw=none,minimum size =0.0em] {} node [below= 0.8cm] {} ; \&
	\node (nEncoder) [BlocksStyle,minimum height= 1.cm] {\Large  Encoder $E_{L-1}$} node [below= 0.8cm]{}; \&
	\node (nChannel) [draw,circle,minimum size=0.3cm,inner sep=0.2em,scale=0.5] {\Large  ${(1+i)^{L-1}}$} node [below= 0.8cm]{}; \&[-4ex]
	\node (nSP) {}; \&[-3ex]
	\node (nDecoder) {
}  node [below= 0.8cm] {}; \&[-1ex] 
\\  \node (bits1) {}; \&
	\node (nInput1)  [BlocksStyle,minimum height= 2.2cm] {\Large  S/P} node [below= 0.8cm] {} ; \&
	\node (nEncoder1) [BlocksStyle,minimum width= 3.0cm,minimum height= 1.cm] {\Large  Encoder $E_{1}$} node [below= 0.8cm]{}; \&
	\node (nChannel1) [draw,circle,minimum size=0.3cm,inner sep=0.2em,scale=0.9] {\Large ${1+i}$} node [below= 0.8cm]{}; \&
	\node (nSP1) [draw,circle,minimum size=0.3cm] {$+$} node [below= 0.8cm] {} node [below= 0.8cm]{}; \&
	\node (nDecoder1) [rectangle,draw,align=center,scale=0.8,minimum height= 2.2cm] {\Large  mod $L$\\ \Large  and offset}  node [below= 0.8cm] {}; 
\&  
	\node (nOutput1) {} node [below] {} ;
\\ \node (bits2) {}; \&
	\node (nInput2)  [draw=none] {} node [below= 0.8cm] {} ; \&
	\node (nEncoder2) [BlocksStyle,minimum width= 3.0cm,minimum height= 1.cm] {\Large  Encoder $E_{0}$} node [below= 0.8cm]{}; \&
	\node (nChannel2) {} node [below= 0.8cm]{}; \&
	\node (nSP2)  {} node [below= 0.8cm]{}; \& 
	\node (nDecoder2) {
}  node [below= 0.8cm] {}; \& 
\\ 
%
};
\draw [LinesStyleb] (bits1) -- node [left,pos=-0.2] {\Large \textbf{b}} (nInput1);
\draw [LinesStyleb] (nInput.center) -- node [above] {\Large $\textbf{b}_{L-1}$}  node [below=0.5cm] {$\bf{\cdots}$} (nEncoder);
\draw [LinesStyleb] (nEncoder) -- node [above] {\Large $\textbf{c}_{L-1}$} node [below=0.5cm] {$\bf{\cdots}$} (nChannel) ;
\draw [LinesStyle] (nSP.center) -- (nChannel);
\draw [LinesStyleb] (nSP.center) -- (nSP1);
\draw [LinesStyle] (nInput1.north) -- (nInput.center);
\draw [LinesStyle] (nInput1.south) -- (nInput2.center);
\draw [LinesStyleb] (nInput1.east) --  node [above] { \Large $\textbf{b}_{1}$} (nEncoder1);
\draw [LinesStyleb] (nEncoder1) -- node [above] {\Large  $\textbf{c}_{1}$} (nChannel1) ;
\draw [LinesStyleb] (nChannel1) -- (nSP1);
\draw [LinesStyleb] (nSP1) -- (nDecoder1);
\draw [LinesStyleb] (nDecoder1) -- node [above] {\Large  $\bf{x}$} node [below=0.1cm,pos=1.1,align=center] {\Large  $2^{L}$-QAM \\ \Large symbols} (nOutput1);
\draw [LinesStyleb] (nInput2.center) -- node [above] {\Large  $\textbf{b}_{0}$} (nEncoder2);
\draw [LinesStyle] (nEncoder2) -- node [above,pos=0.15] { \Large $\textbf{c}_{0}$} (nSP2.center);
\draw [LinesStyleb] (nSP2.center) -- (nSP1);
\end{tikzpicture}}
\caption{Transmitter scheme for MLCM.}
\label{fig:tx_side}
\end{center}
\end{figure} 

%% file: multi_level.tex

We depict in Fig.~\ref{fig:tx_side} the overall structure of the transmitter side. A stream of $K$ information bits $\textbf{b} \triangleq [b_0,\cdots, b_{K-1}]$ is split into $L$ streams, representing the number of levels in the \emph{multi-level coding} scheme.  Each of these $L$ streams contains an amount of bits that is in general different among levels, indicated as $\{K_0, \dots, K_{L-1}\}$. For each level $l$, the length-$K_l$ information bit stream is encoded into a codeword of length $N$ by encoder $E_l$, independently of the other streams. We elaborate on the specific algorithm used for the code rate design and on the choice of the channel encoders $E_l$'s in Sec.~\ref{sec:code_rate_design}. 

\begin{figure}[thb]
\centering
		\includegraphics[scale =0.65]{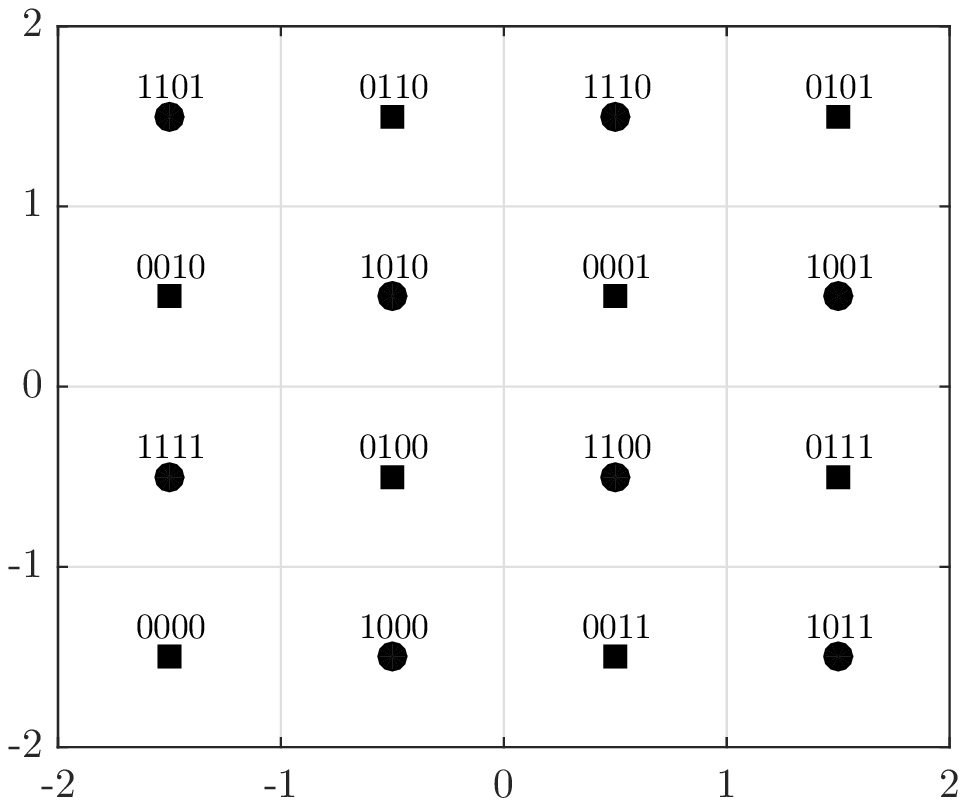}\vspace{-3ex}\caption{Constellation and bit mapping generated by MLCM.}\label{fig:constellation}
\end{figure}

All resulting codewords $\mat{c}_l$, with $l =0, \dots, L-1$, are then multiplied by the $l$-th power of the Gaussian integer $(1+i)$ and added, thus resulting into the vector of complex numbers, $\bar{\mat{x}} = \sum_{l = 0}^{L-1} \mat{c}_l (1+i)^l$. A modulo-$L$ operation over the real and imaginary parts, together with an offset translation are then applied in order to transform each symbol of the vector $\bar{\mat{x}}$ into a symbol of the $2^L$-QAM constellation. An example with $L = 4$, i.e., a 16-QAM constellation, is represented in Fig.~\ref{fig:constellation}. Note that the bit-symbol mapping is different from the BICM-optimal Gray mapping: here symbols corresponding to hypothesis $c_l = 0$ and $c_l = 1$ are as distant as possible, thus improving the decoding performance in a SISO scenario for AWGN channels \cite{mlc99}. In fact for this particular example, the mapping corresponds to the Ungerboeck set-partitioning rule mapping \cite{ung82}. We represent with different shapes and colors the symbols corresponding to values of $c_l = 0$ and $c_l =1$. For example, constellation points represented with a square shape have the first bit $c_0 = 0$, whilst the constellation point has circles for a first bit $c_0 = 1$.   

%% file: code_rate_design.tex
In this section, we describe the MLCM design that we use throughout the paper. The sequence of $K$ information bits $[b_0,\cdots, b_{K-1}]$ is split into $L$ streams of lengths $\{K_0, \dots, K_{L-1}\}$ such that $\sum_{l=0}^{L-1} K_l = K$. 
Each stream $l$ of $K_l$ information bits is encoded using a binary channel code of rate $R_l \triangleq K_l/N$, i.e., each of the $L$ streams is encoded individually with a code of length $N$ and rate $R_l$, such that $\sum_{i=0}^{L-1} K_l/N = K/N = R$. 

Several designs of the code rates have been proposed in \cite{mlc99}, according to different rules, e.g. capacity rule, balanced distance rule or coding exponent rule. In our work, we choose the capacity rule to design the code rates, as explained in the following.

\subsubsection{Capacity Rule}
Let $\mat{y} = \mat{x}+\mat{w}$ be the received signal, $\mat{y} \in \mathbb{C}^N$. In this model, the received signal is simply the superposition of the transmitted signal, $\mat{x}$, with independently drawn and equally probable symbols of a $2^L$-QAM constellation, and the noise at the receiving antenna, which is Gaussian white and circularly symmetric, $\mat{w} \sim \mathcal{CN}(0,\sigma^2 \mat{I}_{N})$. We denote the overall $2^L$-QAM constellation as $\mathcal{S}_0$. 

Given a transmitted point $\mat{x} \in \mathcal{S}_{0}$ corresponding to the vector of coded bits $\textbf{c}=[c_0, \cdots, c_{L-1}]$, the chain rule of mutual information gives
\begin{align}
I(x;y) &= I([c_0, \cdots, c_{L-1}];y)\label{eq:capacity_bit_level} \\
&\hspace{-2ex}= I(c_0;y)+I(c_1;y|c_0)\cdots+I(c_{L-1};y|c_0 \cdots c_{L-2}) \\
&\hspace{-2ex}= \sum_{l} I(c_l;y|c_0 \cdots c_{l-1}) \triangleq \sum_{l} C_l,
\end{align}
where $C_l = I(c_l;y|c_0 \cdots c_{l-1})$ is called, with a slight abuse of language, the bit-level capacity at level $l$. The overall channel capacity is then given by the sum of bit-level capacities, as originally shown in \cite{mlc99}.

\begin{figure}
\centering
\includegraphics[scale=0.48]{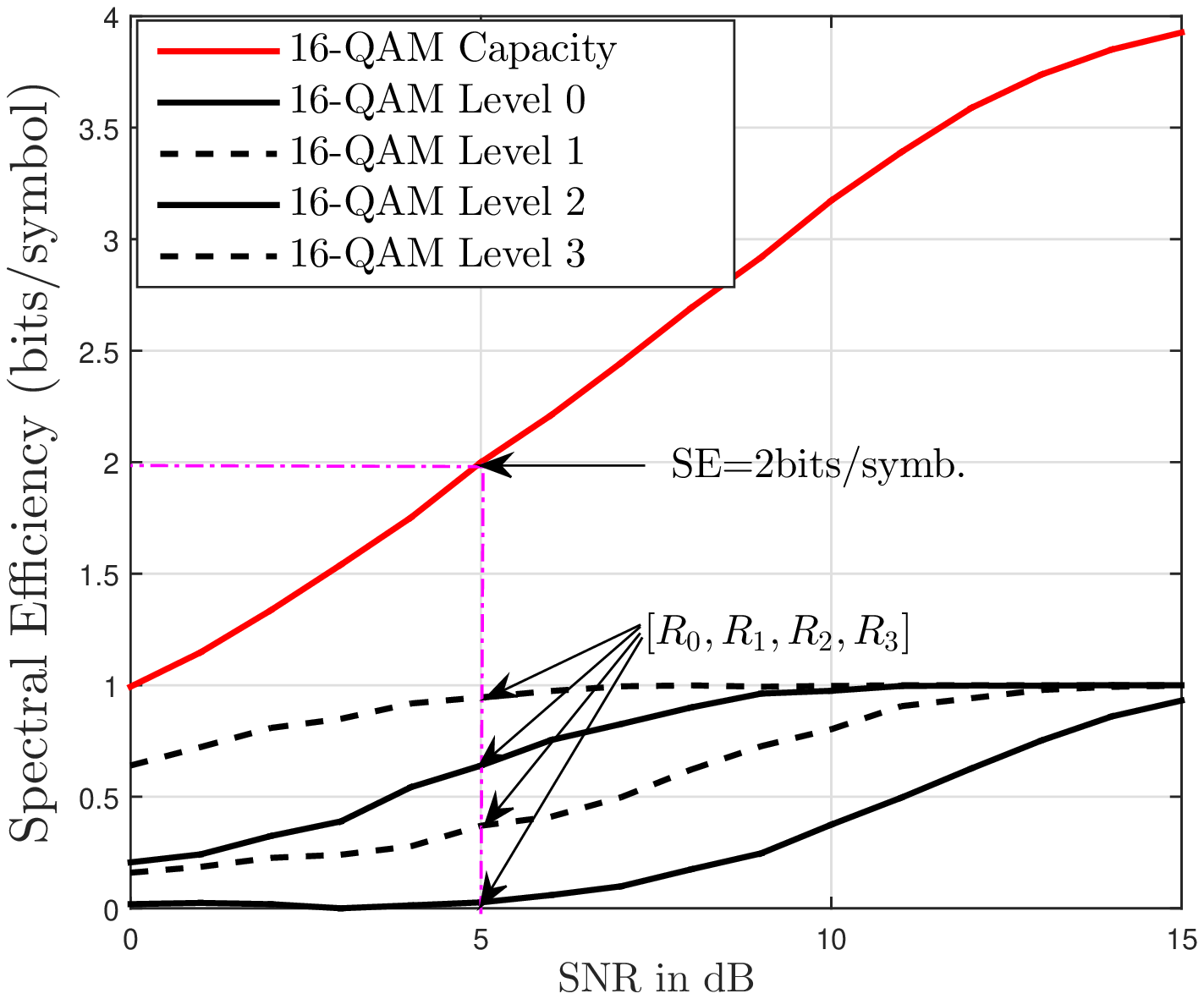}
\vspace{-0.8ex}
\caption{Rate design for SE$=2$bits/symbol with 16-QAM.}\label{fig:levels}
\end{figure}

The capacity-rule design suggests that the rate $R_l$ of the channel code at the individual coding level $l$ of a multi-level coding scheme should be chosen equal to the capacity of the equivalent channel, $C_l$, which would theoretically insure, under multi-stage decoding, that the overall channel capacity is achieved. In order to illustrate the principle, we show in Fig.~\ref{fig:levels} the capacity-rule design for a 16-QAM constellation and an overall target rate of $R=2$ bits/symbol. The curves corresponding to each bit-level capacities are plotted, from which we can obtain the rates $R_0,R_1,R_2,R_3$. The number $N R_l$ is then the number of information bits to be allocated to each individual level, $l$.

We should note that in practice we relax the capacity rule constraint since we are using finite-length channel codes whilst the bit-level capacities results are achieved with infinite codewords. In particular we run Monte-Carlo simulations to guarantee that the chosen rate profile achieves the target bit rate\footnote{A given target bit rate is considered achieved if the Frame Error Rate is smaller than a given threshold, set in our simulations to $10^{-2}$.} with finite codewords. In the following section, we describe the chosen channel coding scheme, namely polar coding \cite{polar}.

\subsubsection{Polar Codes for MLCM}
\input{polar_MLCM.tex}

%% file: polar_MLCM.tex
Polar codes, recently introduced by Arikan in \cite{polar}, are a new class of channel codes. 
They are provably capacity-achieving over various classes of channels, and they provide excellent error rate performance for practical code lengths. 
In their original construction, polar codes are based on the polarization effect of the Kronecker powers of the kernel matrix
$T_2 = \begin{pmatrix}
     1 & 0 \\ 1 & 1 
\end{pmatrix}$, 
referred to as the binary kernel. 
The generator matrix of a polar code is a sub-matrix of $T_2^{\otimes n}$. The unique structure of polar codes offers promising elements for joint optimization of coding and modulation, see \cite{pcm13, isitMLCM}. Hence polar codes are very attractive for coded modulation
schemes, in addition to their natural advantages. In particular, the use of list decoding techniques \cite{list_decoding} for the successive cancellation decoder further increases the performance of the scheme, potentially in addition to the use of a cyclic redundancy check (CRC).

As observed first in \cite{pcm13}, the multi-level coding approach is very similar to polar coding on a
conceptual level, in fact multi-stage decoding works like successive decoding of polar codes. 
In our design, we use polar codes as the component binary channel codes. At each level $l \in [0, \cdots, L-1]$, we decode the polar code of rate $R_l$, whose output is then fed to the demapper for the next bit level $l+1$.
It is well-known that the polar codes at each level
approach each of these bit level capacities when $N$ increases, hence combining polar codes together with multi-level coding (under multi-stage decoding) achieves the coded modulation capacity $C=\sum_{l} C_l$ \cite{isitMLCM}.

The remaining design step is to select the $N-K_l$ frozen bits of the polar code for each bit-channel $l$ out
of all bit channels according to the desired overall code rate, which is done by combining the technique illustrated in Figure \ref{fig:levels} and usual frozen bits selection techniques, e.g. density evolution (DE) algorithms as in \cite{DE_mori}.

%% file: fig_rec.tex
\begin{figure}[!htb]

\begin{center}
\resizebox{0.5\textwidth}{!}{
\begin{tikzpicture}
[
	xscale	= 1,	
	yscale	= 1,	
]

\matrix
[ampersand replacement=\&,
	row sep		= 0.7cm,
	column sep	= 1.09cm,
]
{
   \node (bits) {}; \&[-3ex]
	\node (nInput)  [draw=none,minimum size =0.0em] {} node [below= 0.8cm] {} ; \&
	\node (nEncoder) [BlocksStyle,minimum height= 1.5cm,minimum width= 3.0cm,align=center] {\Large  MPA \\ \Large Stage $L-1$} node [below= 1.1cm]{\Large $\cdots$}; \&
	\node (nChannel) [draw=none] {} node [below= 0.8cm]{}; \&
	\node (nSP) [BlocksStyle,minimum height= 1.5cm,align=center] {\Large  Decoder $D_{L-1}$}; \&[+8ex]
	\node (nDecoder) {
}  node [below= 0.8cm] {}; \&
	\node (nOutput) [draw=none]  {} ;
	\\ \node (bits1) {}; \&
	\node (nInput1)  [draw=none] {} node [below= 0.8cm] {} ; \&
	\node (nEncoder1) [draw=none] {} node [below= 0.8cm]{}; \&
	\node (nChannel1) [BlocksStyle,minimum height= 1.5cm,align=center] {\Large  Encoder $E_{1}$} node [below= 0.8cm]{}; \&
	\node (nSP1)  {} node [below= 0.8cm]{}; \& 
	\node (nDecoder1) {
}  node [below= 0.8cm] {}; \& 
\\  \node (bits2) {}; \&
	\node (nInput2)  [] {} node [below= 0.8cm] {} ; \&
	\node (nEncoder2) [BlocksStyle,minimum height= 1.5cm,minimum width= 3.0cm,align=center] {\Large  MPA \\ \Large Stage $1$} node [below= 0.8cm]{}; \&
	\node (nChannel2) [draw=none] {} node [below= 0.8cm]{}; \&
	\node (nSP2) [BlocksStyle,minimum height= 1.5cm,align=center] {\Large  Decoder $D_{1}$} node [below= 0.8cm] {} node [below= 0.8cm]{}; \&
	\node (nDecoder2) [draw=none] {}  node [below= 0.8cm] {}; 
\&  
	\node (nOutput2) {} node [below] {} ;
\\ 
 \node (bits3) {}; \&
	\node (nInput3)  [draw=none] {} node [below= 0.8cm] {} ; \&
	\node (nEncoder3) [draw=none] {} node [below= 0.8cm]{}; \&
	\node (nChannel3) [BlocksStyle,minimum height= 1.5cm,align=center] {\Large  Encoder $E_{0}$} node [below= 0.8cm]{}; \&
	\node (nSP3)  {} node [below= 0.8cm]{}; \& 
	\node (nDecoder3) {
}  node [below= 0.8cm] {}; \& 
\\ 
 \node (bits4) {}; \&
	\node (nInput4)  [draw=none] {} node [below= 0.8cm] {} ; \&
	\node (nEncoder4)[BlocksStyle,minimum height= 1.5cm,minimum width= 3.0cm,align=center] {\Large  MPA \\ \Large Stage $0$} node [below= 0.8cm]{}; \&
	\node (nChannel4) {} node [below= 0.8cm]{}; \&
	\node (nSP4) [BlocksStyle,minimum height= 1.5cm,align=center] {\Large  Decoder $D_{0}$} node [below= 0.8cm]{}; \& 
	\node (nDecoder4) {
}  node [below= 0.8cm] {}; \&  \node (nOutput4) {} node [below] {} ;
\\ 
%
};
\draw [LinesStyleb] (bits2) -- node [left,pos=-0.2] {\Large \textbf{y}} (nEncoder2);
\draw [LinesStyleb] (nInput.center) -- node [above] {}  node [below=0.5cm] {} (nEncoder);
\draw [LinesStyleb] (nInput4.center) -- node [above] {}  node [below=0.5cm] {} (nEncoder4);
\draw [LinesStyleb] ($(nEncoder.east) + (0,0.2cm)$) -- node [above] {\Large $\textbf{c}_{L-1}(u_6)$} node [below=0.1cm] {$\bf{\cdots}$} ($(nSP.west) + (0,0.2cm)$) ;
\draw [LinesStyleb] ($(nEncoder.east) + (0,-0.2cm)$) -- node [below] {\Large $\textbf{c}_{L-1}(u_1)$}  ($(nSP.west) + (0,-0.2cm)$) ;
\draw [LinesStyleb] ($(nEncoder2.east) + (0,0.2cm)$) -- node [above] {\Large $\textbf{c}_{1}(u_6)$} node [below=0.1cm] {$\bf{\cdots}$} ($(nSP2.west) + (0,0.2cm)$) ;
\draw [LinesStyleb] ($(nEncoder2.east) + (0,-0.2cm)$) -- node [below] {\Large $\textbf{c}_{1}(u_1)$}  ($(nSP2.west) + (0,-0.2cm)$) ;
\draw [LinesStyleb] ($(nEncoder4.east) + (0,0.2cm)$) -- node [above] {\Large $\textbf{c}_{0}(u_6)$} node [below=0.1cm] {$\bf{\cdots}$} ($(nSP4.west) + (0,0.2cm)$) ;
\draw [LinesStyleb] ($(nEncoder4.east) + (0,-0.2cm)$) -- node [below] {\Large $\textbf{c}_{0}(u_1)$}  ($(nSP4.west) + (0,-0.2cm)$) ;
\draw [LinesStyleb] ($(nSP.east) + (0,0.2cm)$) -- node [above] {\Large $\textbf{b}_{L-1}(u_6)$} node [below=0.1cm] {$\bf{\cdots}$} ($(nOutput.east) + (0,0.2cm)$);
\draw [LinesStyleb] ($(nSP.east) + (0,-0.2cm)$) -- node [below] {\Large $\textbf{b}_{L-1}(u_1)$} ($(nOutput.east) + (0,-0.2cm)$);
\draw [LinesStyleb] ($(nSP2.east) + (0,0.2cm)$) -- node [above] {\Large $\textbf{b}_{1}(u_6)$} node [below=0.1cm] {$\bf{\cdots}$} ($(nOutput2.east) + (0,0.2cm)$);
\draw [LinesStyleb] ($(nSP2.east) + (0,-0.2cm)$) -- node [below] {\Large $\textbf{b}_{1}(u_1)$} ($(nOutput2.east) + (0,-0.2cm)$);
\draw [LinesStyleb] ($(nSP4.east) + (0,0.2cm)$) -- node [above] {\Large $\textbf{b}_{0}(u_6)$} node [below=0.1cm] {$\bf{\cdots}$} ($(nOutput4.east) + (0,0.2cm)$);
\draw [LinesStyleb] ($(nSP4.east) + (0,-0.2cm)$) -- node [below] {\Large $\textbf{b}_{0}(u_1)$} ($(nOutput4.east) + (0,-0.2cm)$);
\draw [LinesStyle] (nInput4.center) -- (nInput.center);
\draw [LinesStyle] ($(nDecoder4.east) + (0,-0.2cm)$) -- ($(nDecoder3.east) + (0,-0.cm)$);
\draw [LinesStyleb] ($(nDecoder3.east) + (0,-0.cm)$) -- ($(nChannel3.east) + (0,-0.cm)$);
\draw [LinesStyle] ($(nChannel3.west) + (0,-0.cm)$) -- ($(nEncoder3) + (0,-0.cm)$);
\draw [LinesStyleb] ($(nEncoder3) + (0,-0.cm)$) -- ($(nEncoder2.south) + (0,-0.cm)$);
\draw [LinesStyle] ($(nDecoder2.east) + (0,-0.2cm)$) -- ($(nDecoder1.east) + (0,-0.cm)$);
\draw [LinesStyleb] ($(nDecoder1.east) + (0,-0.cm)$) -- ($(nChannel1.east) + (0,-0.cm)$);
\draw [LinesStyle] ($(nChannel1.west) + (0,-0.cm)$) -- ($(nEncoder1) + (0,-0.cm)$);
\draw [LinesStyleb] ($(nEncoder1) + (0,-0.cm)$) -- ($(nEncoder.south) + (0,-0.8cm)$);
\end{tikzpicture}}
\caption{Receiver scheme for joint MU detection and multi-stage decoding.}
\label{fig:rx_side}
\end{center}
\end{figure} 